\documentclass[sigconf, nonacm]{acmart}
\usepackage{CJKutf8}
\AtBeginDocument{%
  \providecommand\BibTeX{{%
    \normalfont B\kern-0.5em{\scshape i\kern-0.25em b}\kern-0.8em\TeX}}}
\usepackage{multirow}

\usepackage{array}
\usepackage{booktabs}
\usepackage{setspace}
\usepackage{algorithm}
\usepackage{algpseudocode}
\usepackage{multirow}
\usepackage{bm}
\begin{document}
\begin{CJK}{UTF8}{gbsn}

\author{Yijiang Lian}
\email{lianyijiang@baidu.com}
\affiliation{%
  \institution{Baidu}
}
\author{Shuang Li}
\email{lishuang15@baidu.com}
\affiliation{%
  \institution{Baidu}
}
\author{Chaobing Feng}
\email{fengchaobing@baidu.com}
\affiliation{%
  \institution{Baidu}
}
\author{Yanfeng Zhu}
\email{zhuyanfeng@baidu.com}
\affiliation{%
  \institution{Baidu}
}

\title{Quotient Space-Based Keyword Retrieval in Sponsored Search}




\begin{abstract}
  Synonymous keyword retrieval has become an important problem for sponsored search ever since major search engines
  relax the \emph{exact match} product's matching requirement to a synonymous level.
  Since the synonymous relations between
  queries and keywords are quite scarce,
  the traditional information retrieval framework is inefficient in this scenario.
  In this paper, we propose a novel quotient space-based retrieval framework to address this problem.
  Considering the synonymy among keywords
  as a mathematical \emph{equivalence relation}, we can compress the synonymous keywords
  into one representative, and the corresponding quotient space
  would greatly \;reduce \;the \;size \;of \;the \;keyword \;repository.
  \;Then \;an embedding-based retrieval is directly conducted between queries and the keyword representatives.
  To mitigate the semantic gap of the quotient space-based retrieval,
  a single semantic siamese model is utilized to detect both the keyword--keyword and query-keyword synonymous relations.
  The experiments show that with our quotient space-based retrieval method, the synonymous keyword retrieving performance can be greatly improved in terms of memory cost and recall efficiency.
  This method has been successfully implemented in Baidu's online sponsored search system and has yielded a significant improvement in revenue.

\end{abstract}


\keywords{Keyword Compressing, Quotient Space, Retrieval Model, Keyword Matching, Semantic Siamese Model}

\maketitle
\section{Introduction}\label{sec:Introduction}

Sponsored search advertising is one of
the most popular advertising approaches because it can directly target the user's intentions. To match the user's queries with the advertiser's bidded keywords (Keyword in this paper is particularly used to denote queries purchased by the advertisers) plays a significant role in the sponsored search system.
In general, three \emph{match types} are supported:
\emph{exact match}, \emph{phrase match}, and \emph{broad match} \footnote{https://support.google.com/google-ads/answer/7478529?hl=en}.
Among them, \emph{exact match} indicates that
the ads are eligible to appear when a user searches
for the specific keyword or its synonymous variants.

In this paper, we focus on the synonymous keyword matching problem under the \emph{exact match}.
For a given query and a keyword repository (a snapshot of all the keywords committed by the advertisers), the objective is to retrieve as many synonymous keywords as possible. This industrial problem has several unique challenges.
The first challenge is the extremely high precision (>=95\%) required by this commercial product.
The second challenge is that compared with the large number of candidate keywords,  the synonymous relations between queries and keywords are rare, which makes the traditional retrieval framework (e.g., the inverted term-doc link list)  quite inefficient in this scenario, as most of the retrieved  candidates do not satisfy the synonymous requirement.
The third one is the stubborn semantic gap problem in natural language processing.
And the last challenge is the extremely low latency required in an industrial sponsored search system. 

Though the synonymous keywords for an ad-hoc query are sparse compared with the entire keyword repository, synonymous relations among keywords are quite common.
The advertisers tend to enumerate the keyword variations to capture more similar flows.
As a matter of fact, more than one thousand literally different but synonymous keywords of \emph{the price of double eyelid surgery} can be found in Baidu's keyword repository, e.g., \emph{how much is double eyelid surgery}, \emph{how much does it cost to do a double eyelid surgery}.
Our motivation is to use the synonymous equivalence relation to compress the huge keyword repository
into a small quotient set, with which the query--keyword matching can be efficiently conducted between queries and the quotient space representatives.

\begin{figure}[ht]
\centering
\resizebox{.8\totalheight}{!}{
    \includegraphics[width=0.8\textwidth]{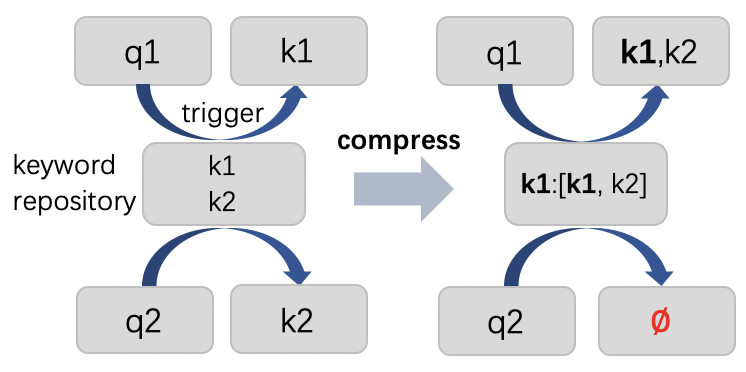}
}
\caption{Semantic gap challenge in quotient-based retrieval.
}
\label{fig:hard_case}

\end{figure}

As is illustrated in Figure \ref{fig:hard_case}, the quotient retrieval puts forward a higher request for overcoming the \emph{semantic gap} challenge.
Suppose that there are two queries Q1, Q2, and two keywords K1, K2,
and the existing synonym matching system can trigger K1 for Q1, and K2 for Q2 separately.
If K2 and K1 are synonymous and compressed into one
representative K1, then Q1 can be matched to \{K1, K2\} effortlessly. However, Q2 might retrieve nothing because of the semantic gap between K1 and Q2.
An intuition to mitigate this problem is to keep the synonym retrieval capability on the (\textit{query},\textit{ keyword}) pairs
synchronized with that on the (\textit{keyword},\textit{ representative}) pairs.

Based on this consideration, we propose the following quotient space-based retrieval approach:
Firstly, a semantic siamese model (SSM) \cite{chicco2021siamese} is trained to calculate the synonymy matching
distance between queries and keywords, as well as between keywords and keywords.
Secondly, approximate nearest neighbor (ANN) search based on the SSM is conducted offline to discover the synonymous relations between keywords. Then, a quotient space, namely a partition of the original keywords, is built based on the synonymous relations, and each partition is denoted by a representative keyword.
Finally, when an ad-hoc query arrives, an ANN search is performed online between the query and the keyword representatives, and all the synonymous keywords in the same quotient space are supplied.

We have successfully applied the proposed method in Baidu's sponsored search system. Experimental results show that using the proposed method, the compression ratio of the keyword space can reach a point of 5:1, the retrieval efficiency is improved, and significant revenue has yielded.

\section{Method}
\subsection{Retrieval Framework}
\begin{figure}
	\centering
	\includegraphics[width=3.3in]{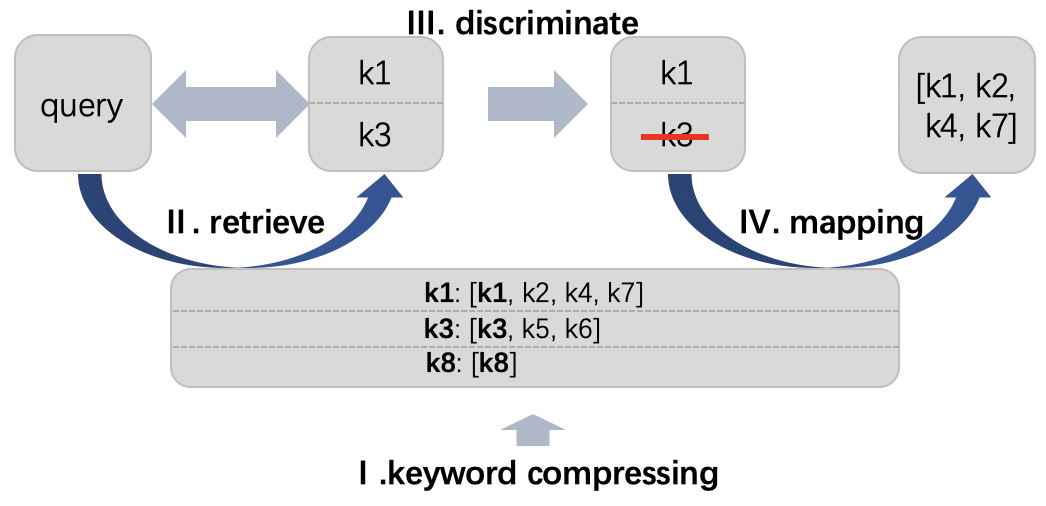}
	\caption{Quotient space-based keyword retrieval framework. \textbf{(Ⅰ)} The original  keyword repository $\mathcal{K}$ 
          is compressed into quotient space representatives $\widetilde{\mathcal{K}}$. \textbf{(Ⅱ)} ANN-based retrieval is conducted between the query and $\widetilde{\mathcal{K}}$, yielding candidate representatives k1 and k3. \textbf{(Ⅲ)} The discriminant model screens out bad representative k3. \textbf{(Ⅳ)} All of the  keywords represented by k1 are returned.}
	\label{fig:framework}
\end{figure}

As is shown in Figure \ref{fig:framework}, our quotient space-based keyword retrieval framework comprises four steps:
The first step is \emph{keyword compressing}. Since the synonymous relationship between keywords are reflexive, symmetric and transitive, it conforms to a mathematical equivalence relationship. Based on this fact, synonymous keyword compressing is conducted for the keyword repository $\mathcal{K}$.
And for each partition, one single keyword is selected as its representative. The set of all the representatives is denoted as $\widetilde{\mathcal{K}}$.
The second step is an embedding-based ANN \emph{retrieval}, which is directly performed between a query $q$ and the keyword representatives $\widetilde{\mathcal{K}}$ to obtain synonymous representative candidates.
The third step is \emph{discrimination}, where a single-tower-like discriminant model is utilized to score the query--representative pairs, only synonymous representatives verified by the discriminant model remain. 
And the final step is keyword \emph{mapping}. For the obtained synonymous representatives, all keywords in the synonym partition indexed by the representatives are fetched out.

\vspace{-0.2cm}

\vspace{-0.2cm}


\subsection{Semantic Siamese Model}\label{sec:SSM}
The SSM plays a core role in our framework, which undertakes two missions: keywords compression and matching between queries and the keyword representatives.
It is double-tower-like as in \cite{facebook_ann_2020}.
 And each tower's function is to encode the input sentence into a semantic
embedding vector, and then these output vectors go through a distance function to measure their
synonymous matching degree.
Since there are no interactions at the encoding stage, this structural restriction
makes the SSM slightly weaker than a single-tower-like model with full interactions. Based on this consideration, a single-tower-like model is initially trained as our discriminant model. And it plays as a teacher model to guide the training of the SSM. In this paper, all of the towers are implemented under the transformer \cite{matsubara2020reranking} network structure.

\textbf{Discriminant Model.}\label{sec:discriminant_model}
The discriminant model is built based on the state-of-the-art bert-like ERNIE \cite{ernie2.0}, which has been well pretrained on large-scale corpora with multitask learning. A two-stage fine-tuning is conducted to match our synonymy discriminating task. The first stage is performed on the training data extracted from the weblog, where the data volume is large (more than 200 million) and the label accuracy is approximately 70\%. The second stage training data is human-annotated, which is 95\% accurate and scarce (less than a million). The AUC(area under the curve) of the trained discriminant model is around 97\%, and the recall approximates 75\% under the precision of 95\%.

\textbf{Semantic Siamese Model.}\label{sec:retrieval_model}
To optimize the SSM's retrieval performance, we adopt the triplet loss \cite{xu2020metric}:
\begin{equation}
L = \sum_{i=1}^{N}max(0, D(q^{i}, k_{+}^{i})-D(q^{i}, k_{-}^{i})+m)
\end{equation}
where $(q, k_{+})$ and $(q, k_{-})$ represent positive pairs and negative pairs, respectively, $D(q, k)$ denotes the synonymy distance between $query$, $keyword$ embedding vector and $m$ is the margin used to separate the positive and negative pairs.

The SSM's training is guided by the discriminant model, which comprises three main steps:

\begin{itemize}
\item [1] Large-scale pretraining on confident instances. In this step, the trained discriminant model is used to score the query--keyword pairs shown in the sponsored weblog $\mathcal{D}$, and the label-confident instances are denoted as $\mathcal{D}_{sure}$, thereby only instances whose scores are higher than the prescribed upper bound or less than the lower bound are retained.
Then the SSM is trained on $\mathcal{D}_{sure}$.

\item [2] Self-training. Given a discriminant model, a query set, and a keyword set, this step forms an iteration loop:
  \begin{itemize}
    \item Based on the current SSM, ANN search is conducted between the query set and the keyword set, and top K approximate nearest keywords are obtained.
    \item The discriminant model is used to score these retrieved query--keyword pairs, and the non-synonymous pairs are selected as hard negative examples.
    \item The hard negative examples are fused into the new train set with the previous data $\mathcal{D}_{sure}$, and the SSM is retrained on this newly obtained dataset.
  \end{itemize}

\item [3] Fine-tuning on human-annotated training data.
\end{itemize}

\renewcommand{\algorithmicrequire}{\textbf{Input:}}
\renewcommand{\algorithmicensure}{\textbf{Output:}}
{\centering
	\begin{minipage}{0.98\linewidth}

		\begin{algorithm}[H]
			\caption{Compressing the keywords into representatives}\label{alg: Keyword Compressing}
			\begin{algorithmic}[1]
				\Require Keyword repository $\mathcal{K}$, SSM $M$, discrimination model $D$, and the number of elements retrieved by ANN $K$
				\Ensure Synonym clusters $C$ of $\mathcal{K}$\\
				Obtain embeddings of the keywords in $\mathcal{K}$ by encoder of $M$;\\
				Build ANN index for embedding representation of $\mathcal{K}$;
				\For{each keyword $k$ in $\mathcal{K}$}
				\State Obtain top $K$ relations of $k$ by ANN;
				\State Filter out non-synonymous relations of $k$ through $D$;
				\EndFor
				\\
				Construct connected subgraphs by synonymy, then each subgraph corresponds to a synonym cluster $c$ in $C$;
				\For{each $c$ in $C$}
				\State Select a representative for $c$ by the relation degree of $k$;
				\State Filter out non-synonymous relations between representative and each $k$ in $c$ through $D$;
				\EndFor
			\end{algorithmic}
		\end{algorithm}
	\end{minipage}
	\par
}

\vspace{-0.2cm}

\subsection{Keyword Compressing} \label{sec:Keyword Compressing}
Keywords compressing  might be simply implemented by inferencing all keyword pairs' synonymous relations
based on the discriminant model. However, it is unrealistic as the keyword's volume is extremely large and the discriminant model's inference is quite time-consuming. The embedding-based SSM (i.e., Sec. \ref{sec:retrieval_model}) can be used here to accelerate the process.
In the meanwhile, the consistency between the retrieval model and the keyword compression model can also mitigate the semantic gap between queries and keyword representatives.

\begin{figure}
	\centering
	\includegraphics[width=3in]{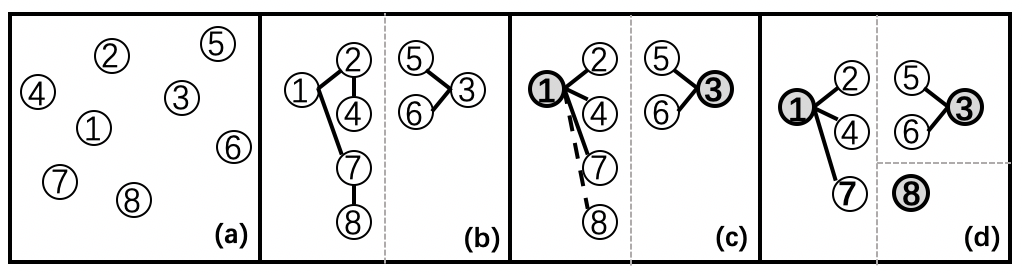}
	\caption{Keyword compressing Procedures: (a) The original keyword repository. (b) \emph{Synonymous relation detection} and \emph{connected subgraph generation}. (c) \emph{Representative selection} and \emph{non-synonymous relation elimination}. (d) Representative cluster generation. (The nodes represent keywords. The grey nodes represent keyword representatives. The solid lines represent synonymous relations, while the dotted lines represent non-synonymous relations.)}
	\label{fig:keyword-compressing}
\end{figure}

As is illustrated in Figure \ref{fig:keyword-compressing}, the keyword compressing consists of three steps.
In the first step (\emph{synonymous relations detection}), ANN is performed between each keyword and the repository to obtain the top synonymous keyword candidates.
The second step (\emph{connected subgraph generation}) is to generate the connected subgraphs based on the detected synonymous relations.
The last step (\emph{representative selection} and \emph{non-synonymous relation elimination}) is to select a representative for each synonym cluster.
In our experiments, the element having the largest relation degree (the number of synonymous relations) is selected to represent the whole cluster. Since the errors might propagate during the subgraph generation, further discrimination is conducted between the representative and all of the other keywords in the same cluster based on the discriminant model. The whole process is elaborated in Algorithm \ref{alg: Keyword Compressing}.

\vspace{-0.2cm}

\vspace{-0.2cm}

\section{Experiments}\label{sec:experiments}

In this section, we focus on the performance of the SSM and the overall performance of the quotient space-based retrieval.

\subsection{Offline Experiments}
\subsubsection{Experiments for the Semantic Siamese Model}
\

\noindent \textbf{Dataset.} A total of 400 thousand query--keyword pairs are randomly sampled from the sponsored search engine's one week's weblog and professionally annotated for their synonymity. This dataset $\mathcal{D}^h$ is further split into three parts with a ratio of 8:1:1, which are used as training ($\mathcal{D}_{train}^h$), developing ($\mathcal{D}_{dev}^h$), and testing data ($\mathcal{D}_{test}^h$) for the SSM respectively.
We focus on two indicators: the Area Under the Curve (AUC) and the recall rate under the precision of 95\% (R@P95 ) on $\mathcal{D}_{test}$.

\textbf{Implementation Details.} Each of the two towers of the SSM is implemented with 4 transformer layers, 4 self-attention heads and the hidden dimension size is 128. These two towers shared their parameters. The discriminant model is implemented with  24 transformer layers, 16 self-attention heads and the hidden dimension size is 1024. During fine-tuning stages, both models are optimized by Adam \cite{kingma2014adam} with an initial learning rate of $1 \times 10^{-4}$ and a batch size of 128.

\textbf{Results.} We conduct an ablation study on the SSM training strategies mentioned in Sec.\ref{sec:retrieval_model}. Four models are compared. All of them go through a final fine-tuning step on $\mathcal{D}_{train}^h$, and their differences are as follows:

\begin{itemize}
\item $M_0$ is pretrained on the raw weblog data $\mathcal{D}$, where \emph{exact-matched} query--keyword pairs are used as positive examples, and the others are considered as negatives;

\item $M_1$ is pretrained on $\mathcal{D}_{sure}$, which is the confident sample subset of $\mathcal{D}$ obtained by the discriminant model;

\item $M_2$ is pretrained on $\mathcal{D}$ and further self-trained based on the discriminant model.

\item $M_3$ is pretrained on $\mathcal{D}_{sure}$ and further self-trained as $M_2$, i.e., $M_3$ implements the full training strategies.
\end{itemize}


\begin{table}[]
	\centering
	\caption{Evaluation of the SSM on $\mathcal{D}_{test}^{h}$. $M_0$ indicates the baseline model pretrained on $\mathcal{D}$. $M_1$ is pretrained on $\mathcal{D}_{sure}$. $M_2$ is pretrained on $\mathcal{D}$ and adopts self-training. $M_3$ applies our full training strategy.}
	\label{tab:Evaluation result on the SynonymousMatch dataset}
	\begin{tabular}{l|cc}
		\hline
		\multicolumn{1}{c|}{} & AUC     & R@P95 \\ \hline
		$M_0$                  & 93.46\% & 43.92\%               \\
		$M_1$           & 94.76\% & 49.60\%               \\
		$M_2$           & 94.16\% & 50.52\%               \\
		\textbf{$M_3$}   & \textbf{95.14\%} & \textbf{55.36\%}               \\ \hline
	\end{tabular}
\end{table}

The results are listed in Table \ref{tab:Evaluation result on the SynonymousMatch dataset}. By comparing $M_1$ with $M_0$, and $M_2$ with $M_0$, we can observe that both the discriminant-model's guidance and the self-training bring positive effects on AUC and R@P95. Besides, $M_3$ gains a further improvement over $M_1$ as well as $M_2$, resulting in 95.14\% on AUC and 55.36\% on R@P95 (+1.68\% on AUC, +11.44\% on R@P95 over $M_0$), which demonstrates the best performance of our full training strategies.



\subsubsection{Experiments for the Quotient Framework}
\

\noindent With a well-trained SSM $M_3$,
we examine the retrieval efficiency of the quotient-based method in this subsection. Our baseline method is the traditional ANN search conducted between queries and the original keyword repository $\mathcal{K}$. The quotient-based method implements the quotient space-based keyword compressing, and conducts ANN search between queries and the keyword representatives $\widetilde{\mathcal{K}}$.

\noindent \textbf{Dataset.}
Our experimental keyword repository $\mathcal{K}$ is a
12 million keyword set sampled from  one day's whole keyword repository, and 10 thousand queries are sampled from one day's weblog. For each query $q$, up to 10 synonymous keywords are obtained by coarse-selection with the discrimination model and refined-selection with human evaluation. This dataset is denoted as $D^{h}_{quo}$. Metric R@K is leveraged to measure the recall rate for $D^h_{quo}$ when top K ANN search is performed, while lantency@K measures the average CPU time cost for conducting the top K ANN search.

\textbf{Implementation Details.}
We adopt HNSW \cite{8594636} both in our offline simulations and online settings for its high efficiency in large-scale index processing. The \emph{number of links for an element} is set to 16, the \emph{size of the dynamic list for the nearest neighbors} is set to 200, and the thread number is set to 10. The test is conducted on
a machine equipped with 28-core  Intel(R) Xeon(R) Gold 5117 CPU clocked at @ 2.00GHz with
a RAM of 250G.

\textbf{Results.} Table \ref{tab:Evaluation result on the KeywordsRetrieval dataset} shows that our quotient-based method can greatly boost recall of the synonymous keywords, where R@10 increases from 31.34\% to 78.75\%, approximately 1.5 times improvement. When top 100 retrieval is conducted, recall rate achieves an increase by 16.23 percentage points.

Table \ref{tab:Time and space consumption on the KeywordsRetrieval dataset}
evaluates the latency and memory cost. The results shows that our quotient-based method can greatly reduce the retrieval latency as well as the memory consumption. Compared with the baseline method, 1/3 latency time for the top 100 retrieval can be saved, and only 23.8\% of the indexing space is consumed.

\vspace{-0.2cm}

\begin{table}[]
	\caption{The recall evaluation of the quotient-based method on $\mathcal{D}^h_{quo}$.}
	\label{tab:Evaluation result on the KeywordsRetrieval dataset}
	\begin{tabular}{l|cc}
		\hline
		 & R@10 & R@100 \\ \hline
		Baseline             & 31.34\%      & 80.27\%       \\
		Quotient-Based      & 78.75\%      & 96.50\%       \\ \hline
	\end{tabular}
\end{table}

\begin{table}[]
	\centering
        \caption{Latency and memory cost test for the quotient-based method. Latency is measured in milliseconds, and memory is measured in gigabytes.}
	\label{tab:Time and space consumption on the KeywordsRetrieval dataset}
	\begin{tabular}{l|cc|c}
		\hline
		 & Latency@10 & Latency@100 & Memory \\ \hline
		Baseline         & 0.26 & 0.90 & 8.4        \\
		Quotient-Based & 0.16 & 0.60 & 2.0        \\ \hline
	\end{tabular}
\end{table}

\subsection{Online Experiments}
This section examines the performance of the overall quotient space-based retrieval framework. The whole process is implemented as the pipeline in Figure \ref{fig:framework} and deployed on Baidu's sponsored search system.
We finally select 460 million active keywords to conduct the keywords compressing, and the ANN index is built upon
80 million keyword representatives correspondingly. Mappers from quotient representatives to keywords are implemented with a lookup table service.
The online A/B experiments show that the quotient space-based retrieval framework leads to a 1.3\% increase in CPM (average revenue per thousand searches),
which is compelling for a large sponsored search system currently. Human evaluations on the shown ads show that the user’s experience is not degraded than before.


\section{Related Work}
\textbf{Paraphrase.} Paraphrases are widely used for various NLP applications, such as document retrieval \cite{2002ACL}, question answering \cite{2003ACl}. Recent works \cite{2018KDD_gan_for_qb, qi2020prophetnetads} proposed generative query--keyword matching methods. \cite{liu2018lcqmc, zhang2019paws,lian2020retrieve} contributed paraphrase datasets for model improvement. Our work is driven by the high precision and efficiency demand in practical industrial application, and dedicates to embedding-based query--keyword synonym retrieval in commercial search scenarios.

\textbf{Embedding Representation and Retrieval.} Embedding representation has been studied and applied in different fields and tasks in the age of deep learning \cite{survey_RepresentationLearning}. \,Bi \,et \,al. \,\cite{embedding_ProductSearch} \,adopted \,a transformer-based embedding retrieval for personalized product search. Huang et al. \cite{facebook_ann_2020} introduced their context-aware embedding representation and retrieval in Facebook vertical searches. As for retrieval efficiency, various ANN algorithms have been proposed. Many of the related works focus on the distribution characteristics of the representation vectors and attempt to reduce the scale of the index, such as principal component analysis (PCA), production quantization \cite{survey_PQ}. Our work differs from the ANN algorithms in that we compress the origin candidates into task-oriented partitions, that is, candidates in a partition are synonymous. Under our retrieval framework, the ANN accelerating algorithms can further work on the retrieval operation smoothly.

\section{Conclusions}
This paper presents a novel quotient space-based retrieval framework
to address the synonymous keyword retrieval problem in sponsored search.
The basic idea is to compress the large keyword space into a small quotient space based on the synonymous equivalence relation, and the keywords matching is directly conducted between the queries and the keyword representatives. A siamese semantic model is trained for embedding-based query--keyword retrieval. Keyword compressing implemented by the same SSM can avoid exacerbating the semantic gap in quotient-based retrieval.
The offline evaluations and real experiments on Baidu's online sponsored search have proved the effectiveness of this work.
This method’s application scope is not limited in the synonymous retrieval under \emph{exact match},
exploration in \emph{phrase match} and \emph{broad match} is also in progress.

\bibliographystyle{ACM-Reference-Format}
\bibliography{ref.bib}


\begin{thebibliography}{17}


\ifx \showCODEN    \undefined \def \showCODEN     #1{\unskip}     \fi
\ifx \showDOI      \undefined \def \showDOI       #1{#1}\fi
\ifx \showISBNx    \undefined \def \showISBNx     #1{\unskip}     \fi
\ifx \showISBNxiii \undefined \def \showISBNxiii  #1{\unskip}     \fi
\ifx \showISSN     \undefined \def \showISSN      #1{\unskip}     \fi
\ifx \showLCCN     \undefined \def \showLCCN      #1{\unskip}     \fi
\ifx \shownote     \undefined \def \shownote      #1{#1}          \fi
\ifx \showarticletitle \undefined \def \showarticletitle #1{#1}   \fi
\ifx \showURL      \undefined \def \showURL       {\relax}        \fi
\providecommand\bibfield[2]{#2}
\providecommand\bibinfo[2]{#2}
\providecommand\natexlab[1]{#1}
\providecommand\showeprint[2][]{arXiv:#2}

\bibitem[\protect\citeauthoryear{{Bengio}, {Courville}, and {Vincent}}{{Bengio}
  et~al\mbox{.}}{2013}]%
        {survey_RepresentationLearning}
\bibfield{author}{\bibinfo{person}{Y. {Bengio}}, \bibinfo{person}{A.
  {Courville}}, {and} \bibinfo{person}{P. {Vincent}}.}
  \bibinfo{year}{2013}\natexlab{}.
\newblock \showarticletitle{Representation Learning: A Review and New
  Perspectives}.
\newblock \bibinfo{journal}{\emph{IEEE Transactions on Pattern Analysis and
  Machine Intelligence}} \bibinfo{volume}{35}, \bibinfo{number}{8}
  (\bibinfo{year}{2013}), \bibinfo{pages}{1798--1828}.
\newblock
\urldef\tempurl%
\url{https://doi.org/10.1109/TPAMI.2013.50}
\showDOI{\tempurl}


\bibitem[\protect\citeauthoryear{Bi, Ai, and Croft}{Bi et~al\mbox{.}}{2020}]%
        {embedding_ProductSearch}
\bibfield{author}{\bibinfo{person}{Keping Bi}, \bibinfo{person}{Qingyao Ai},
  {and} \bibinfo{person}{W.~Bruce Croft}.} \bibinfo{year}{2020}\natexlab{}.
\newblock \bibinfo{booktitle}{\emph{A Transformer-Based Embedding Model for
  Personalized Product Search}}.
\newblock \bibinfo{publisher}{Association for Computing Machinery},
  \bibinfo{address}{New York, NY, USA}, \bibinfo{pages}{1521–1524}.
\newblock
\showISBNx{9781450380164}
\urldef\tempurl%
\url{https://doi.org/10.1145/3397271.3401192}
\showURL{%
\tempurl}


\bibitem[\protect\citeauthoryear{Chicco}{Chicco}{2021}]%
        {chicco2021siamese}
\bibfield{author}{\bibinfo{person}{Davide Chicco}.}
  \bibinfo{year}{2021}\natexlab{}.
\newblock \showarticletitle{Siamese neural networks: An overview}.
\newblock \bibinfo{journal}{\emph{Artificial Neural Networks}}
  (\bibinfo{year}{2021}), \bibinfo{pages}{73--94}.
\newblock


\bibitem[\protect\citeauthoryear{Huang, Sharma, Sun, Xia, Zhang, Pronin,
  Padmanabhan, Ottaviano, and Yang}{Huang et~al\mbox{.}}{2020}]%
        {facebook_ann_2020}
\bibfield{author}{\bibinfo{person}{Jui-Ting Huang}, \bibinfo{person}{Ashish
  Sharma}, \bibinfo{person}{Shuying Sun}, \bibinfo{person}{Li Xia},
  \bibinfo{person}{David Zhang}, \bibinfo{person}{Philip Pronin},
  \bibinfo{person}{Janani Padmanabhan}, \bibinfo{person}{Giuseppe Ottaviano},
  {and} \bibinfo{person}{Linjun Yang}.} \bibinfo{year}{2020}\natexlab{}.
\newblock \showarticletitle{Embedding-Based Retrieval in Facebook Search}. In
  \bibinfo{booktitle}{\emph{Proceedings of the 26th ACM SIGKDD International
  Conference on Knowledge Discovery \& Data Mining}} (Virtual Event, CA, USA)
  \emph{(\bibinfo{series}{KDD '20})}. \bibinfo{publisher}{Association for
  Computing Machinery}, \bibinfo{address}{New York, NY, USA},
  \bibinfo{pages}{2553–2561}.
\newblock
\showISBNx{9781450379984}
\urldef\tempurl%
\url{https://doi.org/10.1145/3394486.3403305}
\showDOI{\tempurl}


\bibitem[\protect\citeauthoryear{Kingma and Ba}{Kingma and Ba}{2015}]%
        {kingma2014adam}
\bibfield{author}{\bibinfo{person}{Diederik~P Kingma} {and}
  \bibinfo{person}{Jimmy Ba}.} \bibinfo{year}{2015}\natexlab{}.
\newblock \showarticletitle{Adam: A method for stochastic optimization}.
\newblock \bibinfo{journal}{\emph{International Conference on Learning
  Representations (ICLR)}} (\bibinfo{year}{2015}).
\newblock


\bibitem[\protect\citeauthoryear{Lee, Gao, and Zhang}{Lee
  et~al\mbox{.}}{2018}]%
        {2018KDD_gan_for_qb}
\bibfield{author}{\bibinfo{person}{Mu-Chu Lee}, \bibinfo{person}{Bin Gao},
  {and} \bibinfo{person}{Ruofei Zhang}.} \bibinfo{year}{2018}\natexlab{}.
\newblock \showarticletitle{Rare Query Expansion Through Generative Adversarial
  Networks in Search Advertising}. In \bibinfo{booktitle}{\emph{Proceedings of
  the 24th ACM SIGKDD International Conference on Knowledge Discovery \& Data
  Mining}} (London, United Kingdom) \emph{(\bibinfo{series}{KDD '18})}.
  \bibinfo{publisher}{Association for Computing Machinery},
  \bibinfo{address}{New York, NY, USA}, \bibinfo{pages}{500–508}.
\newblock
\showISBNx{9781450355520}
\urldef\tempurl%
\url{https://doi.org/10.1145/3219819.3219850}
\showDOI{\tempurl}


\bibitem[\protect\citeauthoryear{Lian, You, Wu, Liu, and Jia}{Lian
  et~al\mbox{.}}{2020}]%
        {lian2020retrieve}
\bibfield{author}{\bibinfo{person}{Yijiang Lian}, \bibinfo{person}{Zhenjun
  You}, \bibinfo{person}{Fan Wu}, \bibinfo{person}{Wenqiang Liu}, {and}
  \bibinfo{person}{Jing Jia}.} \bibinfo{year}{2020}\natexlab{}.
\newblock \bibinfo{title}{Retrieve Synonymous keywords for Frequent Queries in
  Sponsored Search in a Data Augmentation Way}.
\newblock
\newblock
\showeprint[arxiv]{2008.01969}~[cs.IR]


\bibitem[\protect\citeauthoryear{Liu, Chen, Deng, Zeng, Chen, Li, and Tang}{Liu
  et~al\mbox{.}}{2018}]%
        {liu2018lcqmc}
\bibfield{author}{\bibinfo{person}{Xin Liu}, \bibinfo{person}{Qingcai Chen},
  \bibinfo{person}{Chong Deng}, \bibinfo{person}{Huajun Zeng},
  \bibinfo{person}{Jing Chen}, \bibinfo{person}{Dongfang Li}, {and}
  \bibinfo{person}{Buzhou Tang}.} \bibinfo{year}{2018}\natexlab{}.
\newblock \showarticletitle{Lcqmc: A large-scale chinese question matching
  corpus}. In \bibinfo{booktitle}{\emph{Proceedings of the 27th International
  Conference on Computational Linguistics}}. \bibinfo{pages}{1952--1962}.
\newblock


\bibitem[\protect\citeauthoryear{{Malkov} and {Yashunin}}{{Malkov} and
  {Yashunin}}{2020}]%
        {8594636}
\bibfield{author}{\bibinfo{person}{Y.~A. {Malkov}} {and} \bibinfo{person}{D.~A.
  {Yashunin}}.} \bibinfo{year}{2020}\natexlab{}.
\newblock \showarticletitle{Efficient and Robust Approximate Nearest Neighbor
  Search Using Hierarchical Navigable Small World Graphs}.
\newblock \bibinfo{journal}{\emph{IEEE Transactions on Pattern Analysis and
  Machine Intelligence}} \bibinfo{volume}{42}, \bibinfo{number}{4}
  (\bibinfo{year}{2020}), \bibinfo{pages}{824--836}.
\newblock
\urldef\tempurl%
\url{https://doi.org/10.1109/TPAMI.2018.2889473}
\showDOI{\tempurl}


\bibitem[\protect\citeauthoryear{Matsubara, Vu, and Moschitti}{Matsubara
  et~al\mbox{.}}{2020}]%
        {matsubara2020reranking}
\bibfield{author}{\bibinfo{person}{Yoshitomo Matsubara}, \bibinfo{person}{Thuy
  Vu}, {and} \bibinfo{person}{Alessandro Moschitti}.}
  \bibinfo{year}{2020}\natexlab{}.
\newblock \showarticletitle{Reranking for efficient transformer-based answer
  selection}. In \bibinfo{booktitle}{\emph{Proceedings of the 43rd
  International ACM SIGIR Conference on Research and Development in Information
  Retrieval}}. \bibinfo{pages}{1577--1580}.
\newblock


\bibitem[\protect\citeauthoryear{Matsui, Uchida, J{\'e}gou, and Satoh}{Matsui
  et~al\mbox{.}}{2018}]%
        {survey_PQ}
\bibfield{author}{\bibinfo{person}{Yusuke Matsui}, \bibinfo{person}{Yusuke
  Uchida}, \bibinfo{person}{Herv{\'e} J{\'e}gou}, {and}
  \bibinfo{person}{Shin'ichi Satoh}.} \bibinfo{year}{2018}\natexlab{}.
\newblock \showarticletitle{A survey of product quantization}.
\newblock \bibinfo{journal}{\emph{ITE Transactions on Media Technology and
  Applications}} \bibinfo{volume}{6}, \bibinfo{number}{1}
  (\bibinfo{year}{2018}), \bibinfo{pages}{2--10}.
\newblock


\bibitem[\protect\citeauthoryear{Qi, Gong, Yan, Jiao, Shao, Zhang, Li, Duan,
  and Zhou}{Qi et~al\mbox{.}}{2020}]%
        {qi2020prophetnetads}
\bibfield{author}{\bibinfo{person}{Weizhen Qi}, \bibinfo{person}{Yeyun Gong},
  \bibinfo{person}{Yu Yan}, \bibinfo{person}{Jian Jiao}, \bibinfo{person}{Bo
  Shao}, \bibinfo{person}{Ruofei Zhang}, \bibinfo{person}{Houqiang Li},
  \bibinfo{person}{Nan Duan}, {and} \bibinfo{person}{Ming Zhou}.}
  \bibinfo{year}{2020}\natexlab{}.
\newblock \bibinfo{title}{ProphetNet-Ads: A Looking Ahead Strategy for
  Generative Retrieval Models in Sponsored Search Engine}.
\newblock
\newblock
\showeprint[arxiv]{2010.10789}~[cs.IR]


\bibitem[\protect\citeauthoryear{Rinaldi, Dowdall, Kaljurand, Hess, and
  Moll\'a}{Rinaldi et~al\mbox{.}}{2003}]%
        {2003ACl}
\bibfield{author}{\bibinfo{person}{Fabio Rinaldi}, \bibinfo{person}{James
  Dowdall}, \bibinfo{person}{Kaarel Kaljurand}, \bibinfo{person}{Michael Hess},
  {and} \bibinfo{person}{Diego Moll\'a}.} \bibinfo{year}{2003}\natexlab{}.
\newblock \showarticletitle{Exploiting Paraphrases in a Question Answering
  System}. In \bibinfo{booktitle}{\emph{Proceedings of the Second International
  Workshop on Paraphrasing - Volume 16}} (Sapporo, Japan)
  \emph{(\bibinfo{series}{PARAPHRASE '03})}. \bibinfo{publisher}{Association
  for Computational Linguistics}, \bibinfo{address}{USA},
  \bibinfo{pages}{25–32}.
\newblock
\urldef\tempurl%
\url{https://doi.org/10.3115/1118984.1118988}
\showDOI{\tempurl}


\bibitem[\protect\citeauthoryear{Sun, Wang, Li, Feng, Tian, Wu, and Wang}{Sun
  et~al\mbox{.}}{2020}]%
        {ernie2.0}
\bibfield{author}{\bibinfo{person}{Yu Sun}, \bibinfo{person}{Shuohuan Wang},
  \bibinfo{person}{Yukun Li}, \bibinfo{person}{Shikun Feng},
  \bibinfo{person}{Hao Tian}, \bibinfo{person}{Hua Wu}, {and}
  \bibinfo{person}{Haifeng Wang}.} \bibinfo{year}{2020}\natexlab{}.
\newblock \showarticletitle{ERNIE 2.0: A Continual Pre-Training Framework for
  Language Understanding}.
\newblock \bibinfo{journal}{\emph{Proceedings of the AAAI Conference on
  Artificial Intelligence}} \bibinfo{volume}{34}, \bibinfo{number}{05}
  (\bibinfo{date}{Apr.} \bibinfo{year}{2020}), \bibinfo{pages}{8968--8975}.
\newblock
\urldef\tempurl%
\url{https://doi.org/10.1609/aaai.v34i05.6428}
\showDOI{\tempurl}


\bibitem[\protect\citeauthoryear{Xu, Zhang, Cheng, and Chu}{Xu
  et~al\mbox{.}}{2020}]%
        {xu2020metric}
\bibfield{author}{\bibinfo{person}{Furong Xu}, \bibinfo{person}{Wei Zhang},
  \bibinfo{person}{Yuan Cheng}, {and} \bibinfo{person}{Wei Chu}.}
  \bibinfo{year}{2020}\natexlab{}.
\newblock \showarticletitle{Metric Learning with Equidistant and
  Equidistributed Triplet-based Loss for Product Image Search}. In
  \bibinfo{booktitle}{\emph{Proceedings of The Web Conference 2020}}.
  \bibinfo{pages}{57--65}.
\newblock


\bibitem[\protect\citeauthoryear{Zhang, Baldridge, and He}{Zhang
  et~al\mbox{.}}{2019}]%
        {zhang2019paws}
\bibfield{author}{\bibinfo{person}{Yuan Zhang}, \bibinfo{person}{Jason
  Baldridge}, {and} \bibinfo{person}{Luheng He}.}
  \bibinfo{year}{2019}\natexlab{}.
\newblock \showarticletitle{PAWS: Paraphrase Adversaries from Word Scrambling}.
  In \bibinfo{booktitle}{\emph{Proceedings of the 2019 Conference of the North
  American Chapter of the Association for Computational Linguistics: Human
  Language Technologies, Volume 1 (Long and Short Papers)}}.
  \bibinfo{pages}{1298--1308}.
\newblock


\bibitem[\protect\citeauthoryear{Zukerman and Raskutti}{Zukerman and
  Raskutti}{2002}]%
        {2002ACL}
\bibfield{author}{\bibinfo{person}{Ingrid Zukerman} {and}
  \bibinfo{person}{Bhavani Raskutti}.} \bibinfo{year}{2002}\natexlab{}.
\newblock \showarticletitle{Lexical Query Paraphrasing for Document Retrieval}.
  In \bibinfo{booktitle}{\emph{Proceedings of the 19th International Conference
  on Computational Linguistics - Volume 1}} (Taipei, Taiwan)
  \emph{(\bibinfo{series}{COLING '02})}. \bibinfo{publisher}{Association for
  Computational Linguistics}, \bibinfo{address}{USA}, \bibinfo{pages}{1–7}.
\newblock
\urldef\tempurl%
\url{https://doi.org/10.3115/1072228.1072389}
\showDOI{\tempurl}


\end{thebibliography}
\end{CJK}
\end{document}